\documentclass[aps,prd,showpacs,twocolumn]{revtex4}

\RequirePackage[T1]{fontenc}

\usepackage{epsfig}
\usepackage{longtable}

\begin{document}

\title{Spectroscopic Assignments of the Excited $B$-Mesons} 
\author{ Stephen Godfrey\footnote{Email: godfrey@physics.carleton.ca} and Kenneth Moats}
\affiliation{
Ottawa-Carleton Institute for Physics, 
Department of Physics, Carleton University, Ottawa, Canada K1S 5B6 \label{addr1}
}

%

\date{\today}


\begin{abstract}
Excited $B$-mesons have been observed by the D0, CDF, LHCb and CMS experiments.  We use the 
predictions of the relativized quark model to make quark model spectroscopic 
assignments for these states.  We identify the   $B_2^*(5747)$ and $B_1(5721)$
as the $B_2^*[1^3P_2]$ and $B_1[1P_1]$ states and the $B_{s2}^*(5840)$ and $B_{s1}(5830)$
as the $B_{s2}^*[1^3P_2]$ and $B_{s1}[1P_1]$ states.  More information
is needed to identify the  $B_J(5970)$ and $B_J(5840)$ states and we suggest a number
of measurements to make this identification:
the  determination of their $J^P$ quantum numbers and either confirming or ruling out their decays
to the $B\pi$ final state. 
With the current information available we believe it most likely that
the $B_J(5970)$ is the $B^*[2^3S_1]$ state, with the $B_J(5840)$ needing confirmation.
\end{abstract}

\maketitle

\section{Introduction}

Over the past several years a large number of new hadron states have been observed by
various collider experiments \cite{Eichten:2007qx,Olsen:2017bmm,Lebed:2016hpi}.  
While the so-called exotic states  not
conforming to quark model states  have received the bulk of the attention 
\cite{Olsen:2017bmm,Lebed:2016hpi,Godfrey:2008nc,Swanson:2006st,Guo:2017jvc}, 
states that appear to be conventional quark model states can provide a useful test of the
continued utility of the quark model \cite{godfrey85xj,Godfrey:2016nwn,Godfrey:2015dia}.  
Over the last decade the D0 \cite{Abazov:2007af,Abazov:2007vq}, 
CDF \cite{Aaltonen:2008aa,Aaltonen:2007ah,Aaltonen:2013atp},
LHCb \cite{Aaij:2015qla,Aaij:2012uva}
and CMS \cite{Sirunyan:2018grk} 
hadron collider experiments  have observed a number of excited bottom and bottom-strange
mesons.  
We summarize the properties of these states  in Table~\ref{tab:exptproperties}
where we quote the Particle Data Group 
(PDG) values,
averaged over charge states \cite{PDG:2018}. 
At the same time, there have been numerous theoretical calculations of the 
properties of these states 
\cite{Godfrey:2016nwn,Godfrey:1986wj,DiPierro:2001uu,Zhong:2008kd,Ebert:2009ua,Luo:2009wu,Colangelo:2012xi,Xiao:2014ura,Sun:2014wea,Wang:2014cta,Xu:2014mka,Ferretti:2015rsa,Liu:2015lka,Liu:2016efm,Lu:2016bbk,Asghar:2018tha,Gupta:2018xds}.  
In this brief note we will compare the predictions of a 
particular quark model \cite{godfrey85xj,Godfrey:2016nwn} to the measured properties of the recently observed
excited bottom mesons.  We will not describe the model or calculations in any detail  as
these can be found in previous publications \cite{godfrey85xj,Godfrey:2016nwn,Godfrey:2015dia}.  

\begin{table}[t]
\caption{Summary of the observed excited $B$-meson properties.  
Unless the charge state is explicitly labelled, 
the values are Particle Data Group values \cite{PDG:2018} 
averaged over the different charge states.
\label{tab:exptproperties}}
\begin{tabular}{llllll} \hline \hline
State &  $J^P$ & Mass (MeV)  & Width (MeV)  \\
\hline 
$B_1(5721)$   & $1^+$	& $ 5726.0\pm 1.2 $  &   $ 28.5 \pm 3.0 $		\\
\hline
$ B_2^*(5747)$	& $2^+$	& $5738.4 \pm 0.5 $  & $23.8 \pm 1.6  $			\\
			&	&	\multicolumn{2}{l}{ ${{\Gamma (B_2^*(5747)^0 \to B^{*+}\pi^-)}\over{\Gamma (B_2^*(5747)^0 \to B^+\pi^-)}} = 0.82\pm 0.28$ } \\ 
\hline
$B_J(5840) $	& 	& $5860.8\pm 8.1$ & $143.8 \pm 34.6$   \\
\hline
$B_J(5970) $	& 	& $5967.5 \pm 3.5$ & $76.0 \pm 10.3$   \\
\hline
$B_{s1}(5830)$ & $1^+$ & $ 5828.63 \pm 0.27 $   & $ 0.5 \pm 0.4 $   \\
\hline
$B_{s2}^*(5840)$ & $2^+$ & 	$ 5839.85\pm 0.17 $ & $ 1.47\pm 0.33 $   \\
			&	&	\multicolumn{2}{l}{${{\Gamma(B_{s2}^*(5840)^0\to B^{*+}K^-)} \over{\Gamma (B_{s2}^*(5840)^0\to B^+K^-)} } = 0.093\pm 0.018$} \\   
\hline
\hline
\end{tabular}
\end{table}

We begin in Section~\ref{sec:quark_model} with a very brief outline of the quark model and
decay model we are using to make our predictions.  It turns out that the new excited states
fall into natural groupings so we will examine each of these groupings in turn 
and discuss their
spectroscopic assignments.  In these sections we include
two sets of decay results.  The first simply reproduces the results of Ref.~\cite{Godfrey:2016nwn}
which uses the predicted masses in the calculations.  In the second set of results we 
recalculate the decay widths
using the measured masses to properly take into account the phase space. 
In Section~\ref{sec:summary} we summarize our 
results and the suggested measurements that can be used to further discriminate between 
spectroscopic assignments,  in particular for the $B_J(5970)$ and $B_J(5840)$ states.

\section{Bottom Mesons:  Comparison Between Theory and Experiment}
\label{sec:quark_model}

\subsection{A Brief Sketch of the Quark Model}

In a previous publication we presented the results of a comprehensive calculation of 
bottom meson properties \cite{Godfrey:2016nwn}.  We will start by comparing those results to 
the measured properties of the observed excited bottom mesons to associate them with 
specific
quark model states.  Given that the quark model mass predictions do not exactly correspond
with the observed masses we will recompute their decay partial widths using the measured
masses as input to see how this alters the results. 
 
For our predictions, we use the 
relativized quark model \cite{godfrey85xj,Godfrey:2016nwn}.  
It incorporates the colour Coulomb plus linear confining
potential with a running strong coupling constant and relativistic corrections. 
The details of this model can be found in Ref.~\cite{godfrey85xj} 
and \cite{Godfrey:2016nwn,Godfrey:2015dia,Godfrey:1986wj,Godfrey:2005ww,Godfrey:2015dva,Godfrey:2004ya,Godfrey:1985by,godfrey85b} 
to which we refer the interested reader.
The parameters of the model, including the 
constituent quark masses, are given in Ref.~\cite{godfrey85xj}.
This model has been reasonably successful in 
describing most known mesons although in recent years an increasing number 
of states have been observed that do not fit into this picture and as such are often referred to 
as ``exotics'' \cite{Olsen:2017bmm,Lebed:2016hpi,Godfrey:2008nc,Swanson:2006st}.
An important limitation of this model is that it is restricted to the $q\bar{q}$ 
sector of the Fock space and does not take into account 
higher-order components that 
can be described by coupled channel effects 
\cite{Eichten:2004uh,Barnes:2007xu,Geiger:1992va}.  As a consequence 
of neglecting these effects
and the crudeness of the relativization procedure we do not 
expect the mass predictions to be accurate to better than $\sim 10-20$~MeV.

For the case of a quark and antiquark of unequal mass, charge conjugation
parity is no longer a good quantum number so that states with different 
total spins but with the same total angular momentum, such as
$^3P_1 -^1P_1$ and $^3D_2-^1D_2$ pairs,  can mix via
the spin orbit interaction or some other mechanism such as mixing via coupled channels.
Consequently, for example, the physical $J=1$ $P$-wave states are linear
combinations of $^3P_1$ and $^1P_1$:
\begin{eqnarray}
\label{eqn:mixing}
P_{1}  & = {^1P_1} \cos\theta_{nP} + {^3P_1} \sin\theta_{nP} \nonumber \\   
P_{1}^{\prime} & =-{^1P_1} \sin\theta_{nP} + {^3P_1} \cos \theta_{nP} 
\end{eqnarray}
where $P\equiv L=1$ designates the relative orbital angular momentum of the $q\bar{q}$ 
pair and the subscript $J=1$ is the total angular momentum including the spin of the $q\bar{q}$ 
pair, which is equal to $L$, with analogous expressions for other values of $L$.  
$\theta_{nL}$ is found by diagonalizing the mass matrix for the antisymmetric
piece of the spin-orbit interaction (which arises for unequal mass quarks and antiquarks)
in the basis of eigenvectors of the $|jm; ls\rangle$ sectors \cite{godfrey85xj}. 
The resulting mixing angles for each sector are given in the caption of the corresponding table 
 (Tables \ref{tab:B1Pproperties}, \ref{tab:Bs1Pproperties} and \ref{tab:B1Dproperties}).
In the heavy quark limit (HQL) in which the heavy quark mass $m_Q\to \infty$, 
the states can be described by the total angular momentum of the
light quark, $j_q$, which couples to the spin of the heavy quark. 
In this limit the state that is mainly spin singlet has $j_q=l+{1\over 2}$
while the state that is mainly spin triplet has $j_q=l-{1\over 2}$ and is labelled with a prime \cite{Cahn:2003cw}.
For $L\neq 0$ this results in two doublets.  The members of 
the $j_q=l+{1\over 2}$ doublet 
have relatively narrow widths while the members of the $j_q=l-{1\over 2}$ 
doublet are relatively
broad.  In the HQL, $\theta_{J}=\tan^{-1}\left(\sqrt{J\over{J+1}}\right)$ \cite{Cahn:2003cw}.
We note that the definition of the mixing angles is fraught 
with ambiguities and one should be extremely careful comparing predictions from different 
papers \cite{barnes}.

We calculate decay widths using the $^3P_0$ quark 
pair creation model 
\cite{Micu:1968mk,Le Yaouanc:1972ae,Ackleh:1996yt,Barnes:2005pb,Blundell:1995ev}.  
The details of the $^3P_0$ quark 
pair creation model along with our conventions are
summarized in Ref.~\cite{Godfrey:2015dia} with the specifics applied to bottom mesons
given in Ref.~\cite{Godfrey:2016nwn}.
The parameters used in our calculation, $\gamma$, the pair creation parameter and $\beta$, 
a universal oscillator parameter for the light mesons,  were found from fits to light meson 
decays \cite{Blundell:1995ev,Close:2005se,Blundell:1996as}. 
The predicted widths
are fairly insensitive to the precise values used for $\beta$ provided $\gamma$ 
is appropriately rescaled.  However $\gamma$ can vary as much as 30\% and still
give reasonable global fits of light meson decay widths \cite{Close:2005se,Blundell:1996as}.
This can result in factors of two variations to predicted widths, both smaller or larger.

The radiative transition widths were calculated using the expressions from 
Refs.~\cite{Kwo88,JDJ,Nov78}, 
which are reproduced in Ref.~\cite{Godfrey:2016nwn}.

The predicted masses, mixing angles and widths for the the $B(1P)$,   $B_s(1P)$, $B(2S)$
and $B(1D)$ multiplets are given in Tables~\ref{tab:B1Pproperties},
\ref{tab:Bs1Pproperties},  \ref{tab:B1Sproperties} and \ref{tab:B1Dproperties}
respectively.  
The masses and widths in column 3, labelled QM, were obtained using the predicted masses from Ref.~\cite{Godfrey:2016nwn}
for both the initial state and for the heavy decay product.  
The masses in column 4, labelled Expt,  are the measured masses and the widths in column 4 
were obtained using the measured masses
for both the initial state and all the decay products.

The excited $B$-mesons fall into natural groupings.  The $B_2^*(5747)$ and $B_1(5721)$   mesons appear
to be consistent with the $B^*_2(1^3P_2)$ and $B_1(1P_1)$ states and the 
$B_{s2}^*(5840)$  and  $B_{s1}(5830)$  mesons with the $B_{s2}^*(1^3P_2)$ and $B_{s1}(1P_1)$ states.  This
has been previously noted in the literature. 
Quark model assignments for the $B_J(5840)$ and $B_J(5970)$ states are not so obviously
apparent.  We will therefore consider each of these pairs of states in turn.

\begin{table}[t]
\caption{Summary of Quark model predictions for the  $B[1P]$ meson properties.  
The two values for the radiative widths and total widths
correspond to the two $B$ charge states as indicated in the table.
The total width does not necessarily equal the sum of the partial
widths listed in the table, as we have not shown decays to final states with small partial widths.
 The column labelled QM uses the predicted masses
for the state and for the heavy decay product \cite{Godfrey:2016nwn} and the column
labelled Expt  shows the widths recalculated using the measured masses. For the
$B_1$ widths we use the calculated mixing angle of  $\theta_{1P}=30.28^\circ$ 
for the $b\bar{q}[1P]$ states \cite{Godfrey:2016nwn}.
\label{tab:B1Pproperties}}
\begin{center}
\begin{tabular}{llrr} \hline \hline
State &   Property   &  \multicolumn{2}{c}{Value (MeV)}  \\
		&			&  (QM) & (Expt) \\
\hline 
$B_2^*[1^3P_2]$  & Mass & 5797 	& 5738.4	\\
					& $\Gamma (B_2^* \to B \gamma) (ub, \; db) $ & 0.4, 0.1 & 0.4, 0.1 \\
					& $\Gamma (B_2^* \to B\pi)$   & 6.2 & 4.6 \\
					& $\Gamma (B_2^* \to B^*\pi)$ & 5.0 & 4.3 \\
					& Total Width $(ub, \; db)$  &  11.7, 11.4 & 9.3, 9.0 \\
$B_1[1P_1] $			& Mass	& 5777	& 5726.0 \\
					& $\Gamma (B_1 \to B \gamma )(ub, \; db) $ & 0.37, 0.11 & 0.11, 0.03 \\
					& $\Gamma (B_1 \to B \gamma )(ub, \; db) $ & 0.1, 0.03 & 0.3, 0.08\\
					& $\Gamma (B_1 \to B^*\pi )$ & 7 & 6.4 \\
					& Total Width $(ub, \; db)$  & 7.3, 6.9 & 6.8, 6.6 \\
$B_1'[1P_1'] $		& Mass   	& 5784	& 5725.5 \\
					& $\Gamma (B_1' \to B^*\pi )$ & 163 & 160 \\
					& Total Width   & 163 & 160  \\
$B_0^*[1^3P_0] $	 & Mass		& 5756 & 5697.4 	\\
					& $\Gamma (B_0^* \to B\pi )$ & 154 & 148 \\
					& Total Width   & 154 & 148  \\	
\hline
\hline
\end{tabular}
\end{center}
\end{table}

\begin{table}[t]
\caption{Summary of Quark model predictions for the $B_s[1P]$ meson properties. 
$\theta_{1P}=39.12^\circ$ for the $b\bar{s}[1P]$ states \cite{Godfrey:2016nwn}.
The decays $B_{s1}^{(\prime)} \to B^*K$ using the predicted masses 
for the $B_{s1}$ and $B^*$ states  are kinematically
forbidden  and are marked with a dash.
See the caption to Table~\ref{tab:B1Pproperties} for further details.
\label{tab:Bs1Pproperties}}
\begin{center}
\begin{tabular}{llrr} \hline \hline
State &   Property   &  \multicolumn{2}{c}{Value (MeV)}  \\
		&			&  (QM) & (Expt) \\
\hline
$B_{s2}^*[1^3P_2]$ &  Mass &  5876  & 5839.4  \\
					& $ \Gamma (B_{s2}^*\to B_s^* \gamma)$ & 0.11 & 0.10 \\
					& $ \Gamma (B_{s2}^*\to B K)$ & 0.66 & 0.57 \\
					& $ \Gamma (B_{s2}^*\to B^* K)$ & 0.008 & 0.05 \\
					& Total Width   & 0.78 & 0.72 \\					
$B_{s1}[1P_1]$ 		&  Mass 	& 5857  & 5828.63   \\
					& $ \Gamma (B_{s1}\to B_s \gamma)$ & 0.07 & 0.05 \\
					& $ \Gamma (B_{s1}\to B_s^* \gamma)$ & 0.04 & 0.06 \\
					& $\Gamma (B_{s1} \to B^* K )$ 		& -	& 0.34 \\
					& Total Width   & 0.11 & 0.45 \\					
$B_{s1}'[1P_1']$ 	&  Mass		& 5861 & 5824.84    \\
					& $ \Gamma (B_{s1}'\to B_s \gamma)$ & 0.05 & 0.07 \\
					& $ \Gamma (B_{s1}'\to B_s^* \gamma)$ & 0.06 & 0.04 \\
					& $\Gamma (B_{s1}' \to B^* K )$ 		& -	& 77.3 \\
					& Total Width   & 0.11 & 77.4\\					
$B_{s0}^*[1^3P_0]$  &  Mass		&	5831  & 5794.84  \\
					& $ \Gamma (B_{s0}^*\to B K)$ & 138 & 128  \\
					& Total Width   & 138 & 128 \\					
\hline
\hline
\end{tabular}
\end{center}
\end{table}

\begin{table}[t]
\caption{Summary of Quark model predictions for the $B[2S]$ meson properties. 
See the caption to Table~\ref{tab:B1Pproperties} for further details.
\label{tab:B1Sproperties}}
\begin{center}
\begin{tabular}{llrr} \hline \hline
State &   Property   &  \multicolumn{2}{c}{Value (MeV)}  \\
		&			&  (QM) & (Expt) \\
\hline
$ B^*[2^3S_1] $ 	& Mass	&  5933  &	5967.5	\\
					& $\Gamma (B^*(2S) \to {B\pi})$ & 36 & 39\\  
					& $\Gamma (B^*(2S)\to{B^*\pi})$ & 68 & 80 \\
					& $\Gamma (B^*(2S) \to{B\eta})$ & 2 & 4\\
					& $\Gamma (B^*(2S) \to{B^*\eta})$ & 0.4 & 6 \\
					& $\Gamma (B^*(2S) \to{B_s K})$ & 2 & 8\\
					& $\Gamma (B^*(2S) \to{B^*_s K})$ & -  & 8 \\
					& Total Width   & 108 & 146 \\
$ B[2^1S_0] $ 		& Mass	&  5904  & 5860.8	\\
					& $\Gamma (B(2S) \to{B^*\pi})$ & 95 & 96\\
					& Total Width   & 95 & 96 \\								
\hline
\hline
\end{tabular}
\end{center}
\end{table}

\begin{table}[t]
\caption{Summary of Quark model predictions for the $B[1D]$ meson properties. 
$\theta_{1D}=39.69^\circ$ for the $b\bar{q}[1D]$ states \cite{Godfrey:2016nwn}.
See the caption to Table~\ref{tab:B1Pproperties} for further details.
\label{tab:B1Dproperties}}
\begin{center}
\begin{tabular}{llrr} \hline \hline
State &   Property   &  \multicolumn{2}{c}{Value (MeV)}  \\
		&			&  (QM) & (Expt) \\
\hline 
$ B^*_1[1^3D_1]  $ & Mass	&  6110  & 5967.5		\\
					& $\Gamma (B_1^* \to B\pi)$ & 60 & 55 \\
					& $\Gamma (B_1^* \to B^*\pi)$ & 30 & 27 \\
					& $\Gamma (B_1^* \to B\eta)$ & 10 & 5\\
					& $\Gamma (B_1^* \to B^*\eta)$ & 4 & 2 \\
					& $\Gamma (B_1^* \to B(1P_1)\pi)$ & 63 & 43\\
					& $\Gamma (B_1^* \to B_sK)$ 		& 19 & 7\\
					& $\Gamma (B_1^* \to B^*_sK)$ 	& 7 & 2\\
					& $\Gamma (B_1^* \to B\rho$ & 2 & - \\
					& Total Width   & 197 & 140\\
$ B_2[1D_2] $ 		& Mass	&  6095  & 5967.5			\\					
					& $\Gamma (B_2 \to B^*\pi)$ & 20 & 9.4 \\
					& $\Gamma (B_2 \to B\rho)$ & 1.6 & - \\
					& Total Width   & 23 &  10 \\								
$ B_2'[1D_2'] $ 	& Mass	&  6124	& 5967.5		\\					
					& $\Gamma (B_2' \to B^*\pi)$ & 96 & 87\\
					& $\Gamma (B_2' \to B^*\eta)$ & 14 & 5.5  \\
					& $\Gamma (B_2' \to B^*(1^3P_2)\pi)$ & 74 & 47 \\
					& $\Gamma (B_2' \to B_s^* K)$ & 27 & 5\\
					& Total Width   & 213 &   145  \\								
$ B^*_3[1^3D_3]  $ & Mass	&  6106 	& 5967.5		\\
					& $\Gamma (B_3^* \to B\pi)$ & 14 & 6 \\
					& $\Gamma (B_3^* \to B^*\pi)$ & 14 & 6\\
					& Total Width   & 31 & 13  \\								
\hline
\hline
\end{tabular}
\end{center}
\end{table}

\subsection{The $B_2^*(5747)$ and $B_1(5721)$ States}

The $B_2^*(5747)$ and $B_1(5721)$ are both relatively narrow \cite{Aaij:2015qla}.  
Their measured $J^P$ 
quantum numbers are those of the $j_q=3/2$ doublet of a heavy-light meson in the heavy quark limit.
Their masses and widths are also roughly consistent with the quark model predictions for these 
states shown in Table~\ref{tab:B1Pproperties}, although the predicted masses are both about 50 MeV higher 
than observed and the predicted widths are smaller than observed.   The high mass prediction
has been observed in other heavy-light systems such as charm mesons so given this pattern
it is reasonable to identify the $B_2^*(5747)$ and $B_1(5721)$ as the $1^3P_2$ and $1P_1$ $B$ 
mesons and deem the inconsistency in masses as a weakness of the model.  The predicted widths
are roughly a factor of two smaller than the observed widths.  As pointed out above, 
this is within the predictive power
of the quark pair creation model.  The important prediction is that the $j_q=3/2$ $P$-wave doublet, 
consisting of the 
$1^3P_2$ and $1P_1$ states,
is predicted to be narrow and the $j_q=1/2$ doublet, consisting of the $1P_1'$ and 
$1^3P_0$ states,
is predicted to be broad.  Interestingly, one of the earliest calculations of these widths using
the flux-tube breaking model (and another using the pseudoscalar 
emission model) 
is in better agreement with the measured widths \cite{Godfrey:1986wj}.

We recalculated the widths  using the  measured $B_2^*$ and $B_1$ masses, which are shown in 
column 4 of Table~\ref{tab:B1Pproperties}.  The $B_1'$ and
$B_0^*$ masses were obtained by subtracting the predicted $B_2^*- B_1'$ and 
$B_2^*- B_0^*$ mass differences from the measured $B_2^*$ mass.  Not surprisingly, there
is no qualitative difference from the results using the predicted masses although the widths
are slightly smaller due to the reduced phase space. We also recalculated the widths for the $1P_1$
states using the HQL mixing angle of $\theta_{1P} = 35.3^\circ$ 
 and although the narrow $B_1$ width changes slightly
it does not alter our conclusion.  

A further test of this assignment is the predicted versus measured ratio of partial widths
to $B^*\pi$ and $B\pi$ final states.  The PDG average for the ratio is  
\cite{PDG:2018}: 
\begin{equation}
{{\Gamma(B_2^{*0}\to B^{*+}\pi^-)}\over{\Gamma(B_2^{*+}\to B^+\pi^-)}}=0.82 \pm 0.28 .
\end{equation}
This is compared to the predicted
ratio: using the predicted $B_2^*$ mass as input we obtain 
$\Gamma(B_2^{*}\to B^{*}\pi) / \Gamma(B_2^*\to B\pi)=0.81$ 
and using the measured
$B_2^*$ mass as input we obtain 
$\Gamma(B_2^{*}\to B^{*}\pi) / \Gamma(B_2^*\to B\pi)=0.93$.
Both values are consistent with the measured value.

\subsection{The $B_{s2}^*(5840)$ and $B_{s1}(5830)$ States}

These states follow a similar pattern to the $P$-wave $B$-mesons of the previous subsection.  The 
$B_{s2}^*$ and $B_{s1}$ states have properties consistent with the $j_q=3/2$ $P$-wave $B_s$ doublet.
The predicted masses are roughly 30-40 MeV higher than the observed masses and the
predicted decay widths are consistent with the measured widths.  It is worth pointing out that
for these states
the radiative widths make a significant contribution to the total width. Using the 
observed masses to calculate the widths does not change these conclusions.  As before, we 
can also compare the ratio of widths to the $B^* K$ and $BK$ final states.  The measured
ratio is \cite{Aaij:2012uva} 
\begin{equation}
{{\Gamma(B_{s2}^*\to B^{*+}K^-)}\over{\Gamma(B_{s2}^*\to B^{+}K^-)}}=0.093\pm 0.018. 
\end{equation}
The predicted ratio is 0.012 using the predicted $B_{s2}^*$ mass as input and 0.09 using
the measured $B_{s2}^*$ mass as input.  The ratio calculated using the measured 
$B_{s2}^*$ mass as input is in good agreement with experiment giving further support
to the identification of the $B_{s2}^*(5840)$ as the $B_{s2}^* (1^3P_2)$ state.

For the $B_{s1}$ state, 
we note that using the HQL mixing angle 
of $\theta_{1P} = 35.3^\circ$, 
 $\Gamma[B_{s1}\to B^* K]$ is reduced from
0.34~MeV to 0.02~MeV and the total width is dominated by the radiative transitions.  Nevertheless,
given the large experimental uncertainty on the total width, the width in the HQL is also
consistent with experiment. Given how close the predicted mixing angle is to the HQL value,
a more precise measurement of the width would be an
interesting constraint on the $^3P_1-^1P_1$ mixing angle.

\subsection{The $B_J(5970)$ and $B_J(5840)$ States}

The identification of the $B_J(5970)$ and $B_J(5840)$ states is less obvious.  LHCb has
suggested that these states can be identified with the $2^3S_1$ and $2^1S_0$ bottom states,
respectively \cite{Aaij:2015qla}.  However, we note that the PDG has omitted the $B_J(5840)$ from the
summary tables, so the experimental situation should be regarded as inconclusive.  With
this caveat, we explore the spectroscopic possibilities for these states. 

The QM predicts the  $2^3S_1 - 2^1S_0$ mass splitting to be
29~MeV versus the measured mass splitting of 107~MeV.  The predicted splitting is consistent
with the expectation that the $2^3S_1 - 2^1S_0$ splitting be smaller than the $1^3S_1 - 1^1S_0$
mass splitting which is measured to be 45~MeV.  So the $B_J(5970)$ and $B_J(5840)$ mass
splitting is a red flag that their identification with the $2^3S_1$ and $2^1S_0$ states
is questionable.  

We start by associating the $B_J(5970)$ with the $2^3S_1$ bottom meson.  The predicted
mass is about 35~MeV below the measured mass.  This is within the predictive reliability of
the model, although it should be noted that the predicted bottom masses are typically above
the experimental values not below.  The total width is calculated 
to be 108~MeV using the
predicted $2^3S_1$ mass, versus the
measured width of $76.0\pm 10.3$~MeV.  Again, these values are consistent within the predictive
power of the model.  If we recalculate the total width using the measured mass as input we
obtain a total width of 146~MeV, again acceptable although at the limits of acceptability.
With this assignment we consider whether the $B_J(5840)$ can be identified with the
$2^1S_0$ $B$-meson.  The predicted $2^1S_0$ $B$-meson mass is 43~MeV above the measured mass and
the predicted total width using the predicted mass as input is 95~MeV versus the measured value 
of $143.8\pm 34.6$~MeV.  Recalculating the width using the measured 
mass as input only changes the value of the width 
slightly, to 96~MeV. In both cases the mass and width are consistent with 
the $2^1S_0$ $B$-meson state within the predictive power of the model and the experimental
uncertainties,  especially for the total width.  However, as pointed out above,  the
mass splitting between the $B_J(5970)$ and $B_J(5840)$ is a red flag that something is amiss.
Further,  both states have been seen to decay to $B^*\pi$ and ``possibly seen'' decaying
to $B\pi$. 
The latter final state, if confirmed, would disallow the $2^1S_0$ 
identification of the $B_J(5840)$ as the decay $B[2^1S_0]\to B\pi$  is forbidden.  Either
confirming this decay or ruling it out would clarify the situation.  Measuring
the ratio 
$\Gamma(B_J \to B\pi)/\Gamma(B_J \to B^*\pi)$, 
which is predicted to be $\sim 0.5$
for the $B[2^3S_1]$ state, would help confirm the identity of these states.  

Because the $B_J(5970) - B_J(5840)$ mass splitting is inconsistent with 
both the predicted and expected $2^3S_1 - 2^1S_0$ mass splittings,
let us consider a second scenario where
we identify the  $B_J(5840)$ as the $B[2^3S_1]$.  Using
its mass of 5860.8~MeV we obtain a total width of 95~MeV, which is consistent, within experimental
uncertainties, with the $B_J(5840)$ width.  
If we identify the $B_J(5840)$
with the $B[2^3S_1]$, then 
 what is the $B_J(5970)$ state?  The states nearest in mass to 
5968~MeV are the $1D$ states. Their properties are summarized in Table~\ref{tab:B1Dproperties}.
The first thing to note is that the predicted $1D$ masses are $\sim 150$~MeV higher than
the $B_J(5970)$  mass.  Let us set this aside for a moment.  The four $1D$ states are grouped
into a $j_q=3/2$ doublet comprised of the $1^3D_1$ and $1D_2'$ states and a $j_q=5/2$ doublet
comprised of the $1D_2$ and  $1^3D_3$ states.  The $j_q=5/2$ states are narrow with their
widths inconsistent with the $B_J(5970)$ width, even taking into account both experimental 
and theoretical uncertainties.  Thus, if we identify the $B_J(5970)$ with a $1D$ state it
must be either the $1^3D_1$ or $1D_2'$ state. 
Although the predicted widths are almost a factor of two larger, 
as stated previously, this is within the uncertainties of the 
predictions.  These two possible assignments can be distinguished by observing decays to the $B\pi$ final state.  
The $1^3D_1$ state has a significant branching ratio (BR) to $B\pi$ of 40\% compared to its
BR to $B^*\pi$ of 19\%, whereas the $1D_2'$ state is forbidden to decay to $B\pi$.

A final  possibility that should be pointed out is that 
the $B_J(5970)$ is the $2^3S_1$ state but that  the $B_J(5840)$
is not confirmed  and the $2^1S_0$ 
state has yet to be observed. 

From the discussion above, it is clear that confirming or ruling out the decays of 
the $B_J(5970)$ and $B_J(5840)$ states  to $B\pi$  is crucial to making
spectroscopic assignments for these states.  
If the $B_J(5840)$ is confirmed to decay
to $B\pi$ it would rule out the $2^1S_0$ assignment while if 
the decay to $B\pi$ can be ruled out with some
confidence, it would support that assignment.  
The latter case makes a strong case
for identifying the $B_J(5840)$ and $B_J(5970)$ as the $B(2^1S_0)$ and $B(2^3S_1)$ states
respectively.  If the $B_J(5840)$ is confirmed to decay to $B\pi$, this assignment is no longer viable and the 
$B_J(5840)$ would be identified as the $B(2^3S_1)$.
 This opens  up the possibility that the
$B_J(5970)$ is a $1D$ state, 
either the $1^3D_1$ if its decay to $B\pi$ is confirmed, 
or the $1D_2'$
otherwise.   In any of these scenarios, determining the $J^P$ quantum numbers is a
crucial piece of the puzzle.

\section{Summary and Conclusions}
\label{sec:summary}

In this paper we reviewed the possible spectroscopic assignments for the recently observed
excited $B$ and $B_s$ mesons.  As pointed out by others, the properties of the
$B_2^*(5747)$ and $B_1(5721)$ are consistent with those of the $B_{2}^*$ and $B_{1}$
quark model states.  Likewise the properties of the $B_{s2}^*(5840)$ and $B_{s1}(5830)$ are
consistent with those of $B_{s2}^*$ and $B_{s1}$ states.  The identification of the
$B_J(5970)$ and $B_J(5840)$ states is problematic and needs further information: the $J^P$ 
quantum numbers, confirmation of the $B_J(5840)$ and confirmation of the $B\pi$ decay mode for both states.  
With the current information the most likely conclusion is that the $B_J(5970)$ is associated
with the $B(2^3S_1)$ state and the $B(2^1S_0)$ has yet to be observed.  If the
$B_J(5840)$ is confirmed with the correct quantum numbers and does not decay to $B\pi$ 
it can be identified as the $B(2^1S_0)$ state.  If it does decay to $B\pi$, it could
be identified as the $B(2^3S_1)$ with the $B(2^1S_0)$  not yet observed and the 
$B_J(5970)$ identified as a $1D$ state, either the $1^3D_1$ or $1D_2'$  depending on whether
or not it is confirmed to decay to $B\pi$.
 We consider the latter explanation unlikely
as we believe the $B_J(5970)$  mass to be inconsistent with a $1D$ state.


\begin{acknowledgements}
This research was supported in part 
by the Natural Sciences and Engineering Research Council of Canada under grant number SAPIN-2016-00041. 
\end{acknowledgements}


\end{document}